\documentclass[twocolumn,english,aps,prd,nofootinbib]{revtex4}
\usepackage{lmodern}
\usepackage[T1]{fontenc}
\usepackage[latin9]{inputenc}
\usepackage{xcolor}
\usepackage{babel}
\usepackage{amsmath}
\usepackage{amsthm}
\usepackage{amssymb}
\usepackage{esint}
\usepackage[unicode=true]
 {hyperref}

\makeatletter
\@ifundefined{textcolor}{}
{%
 \definecolor{BLACK}{gray}{0}
 \definecolor{WHITE}{gray}{1}
 \definecolor{RED}{rgb}{1,0,0}
 \definecolor{GREEN}{rgb}{0,1,0}
 \definecolor{BLUE}{rgb}{0,0,1}
 \definecolor{CYAN}{cmyk}{1,0,0,0}
 \definecolor{MAGENTA}{cmyk}{0,1,0,0}
 \definecolor{YELLOW}{cmyk}{0,0,1,0}
}
\theoremstyle{plain}
\newtheorem{thm}{\protect\theoremname}
\theoremstyle{plain}
\newtheorem{lem}[thm]{\protect\lemmaname}
\ifx\proof\undefined
\newenvironment{proof}[1][\protect\proofname]{\par
	\normalfont\topsep6\p@\@plus6\p@\relax
	\trivlist
	\itemindent\parindent
	\item[\hskip\labelsep\scshape #1]\ignorespaces
}{%
	\endtrivlist\@endpefalse
}
\providecommand{\proofname}{Proof}
\fi
\theoremstyle{remark}
\newtheorem{rem}[thm]{\protect\remarkname}

\usepackage{enumerate}

\makeatother

\providecommand{\lemmaname}{Lemma}
\providecommand{\remarkname}{Remark}
\providecommand{\theoremname}{Theorem}

\begin{document}
\title{Beyond Schwarzschild-de Sitter spacetimes: III. $\ $A perturbative
vacuum with\vskip2pt non-constant scalar curvature in \textmd{$R+R^{2}$}
gravity}
\author{Hoang Ky Nguyen$\,$}
\email[\ \,Email:\ \ ]{hoang.nguyen@ubbcluj.ro}

\affiliation{Department of Physics, Babe\c{s}--Bolyai University, Cluj-Napoca
400084, Romania}
\date{\today}
\begin{abstract}
\vskip2pt In violation of the generalized Lichnerowicz theorem advocated
in \citep{Lu-2015-a,Lu-2015-b,Nelson-2010,Luest-2015-backholes},
quadratic gravity admits vacua with \emph{non-constant} scalar curvature.
In a recent publication \citep{Nguyen-2022-Buchdahl}, we revitalized
a program -- that Buchdahl originated but prematurely abandoned circa
1962 \citep{Buchdahl-1962} -- and uncovered a novel exhaustive class
of static spherically symmetric vacua for pure $\mathcal{R}^{2}$
gravity. The Buchdahl-inspired metrics we obtained therein are \emph{exact}
solutions which exhibit \emph{non-constant} scalar curvature. A product
of fourth-order gravity, the metrics entail a new (Buchdahl) parameter
$k$ which allows the Ricci scalar to vary on the manifold. The metrics
are able to \emph{defeat} the generalized Lichnerowicz theorem by
evading an overly strong restriction on the asymptotic falloff in
the spatial derivatives of the Ricci scalar as assumed in the theorem
\citep{Lu-2015-a,Lu-2015-b,Nelson-2010,Luest-2015-backholes}. The
Buchdahl parameter $k$ is a new characteristic of pure $\mathcal{R}^{2}$
gravity, a higher-derivative theory. By venturing that the Buchdahl
parameter should be a universal hallmark of higher-derivative gravity
at large, in this paper we seek to extend the concept to the quadratic
action $\mathcal{R}^{2}+\gamma\left(\mathcal{R}-2\Lambda\right)$.
We determine that, \emph{up to the order }$\mathcal{O}\left(k^{2}\right)$,
the quadratic field equation admits the following vacuum solution
\begin{align*}
ds^{2} & =e^{k\,\varphi(r)}\left[-\Psi(r)dt^{2}+\frac{dr^{2}}{\Psi(r)}+r^{2}d\Omega^{2}\right]
\end{align*}
in which $\Psi(r):=1-\frac{r_{\text{s}}}{r}-\frac{\Lambda}{3}r^{2}$
and the function $\varphi(r)$ obeys a linear second-order ordinary
differential equation, per
\[
6\left(r^{2}\Psi(r)\varphi'(r)\right)'=\gamma\,r^{2}\varphi(r)\ \ \ \text{subject to }\ \varphi(r\rightarrow\infty)=0
\]
Conforming with our intuition, the Ricci scalar carries the footprint
of a higher-derivative characteristic $k$, given by
\[
\mathcal{R}(r)=4\Lambda-k\left(4\Lambda+\frac{\gamma}{2}\right)\varphi(r)+\mathcal{O}\left(k^{2}\right)
\]
The Ricci scalar is \emph{non-constant}, \emph{including the asymptotically
flat case} (i.e., $\Lambda=0$) as long as $k\neq0$ and $\gamma\neq0$.
The existence of such an asymptotically flat vacuo with non-constant
scalar curvature defeats the generalized Lichnerowicz theorem in \emph{its
entirety}. Our finding thus warrants restoring the $\mathcal{R}^{2}$
term in the full quadratic action, $\gamma\,\mathcal{R}+\beta\,\mathcal{R}^{2}-\alpha\,\mathcal{C}^{\mu\nu\rho\sigma}\mathcal{C}_{\mu\nu\rho\sigma}$,
when applying the L\"u-Perkins-Pope-Stelle ansatz \citep{Lu-2015-a,Lu-2015-b}.
Implications to the L\"u-Perkins-Pope-Stelle solution are discussed
herein.
\end{abstract}
\maketitle

\section{\label{sec:Motivation}Motivation: $\ $Escapes from the\vskip1pt
generalized Lichnerowicz theorem}

In \citep{Nelson-2010} Nelson offered a proof concerning the vacua
of the quadratic action (with $\alpha$, $\beta$, $\gamma$ all non-negative)
\begin{equation}
\gamma\,\mathcal{R}+\beta\,\mathcal{R}^{2}-\alpha\,\mathcal{C}^{\mu\nu\rho\sigma}\mathcal{C}_{\mu\nu\rho\sigma}\label{eq:quadratic-action}
\end{equation}
His proof contains two parts: the trace part and the non-trace part.
The trace part concluded that static vacua of quadratic gravity must
be Ricci-scalar-flat, viz. $\mathcal{R}\equiv0$, if $\gamma>0$.
The non-trace part enforced an even stronger result; it posited that
these vacua must be Ricci-flat, viz. $\mathcal{R}_{\mu\nu}\equiv0$.
In \citep{Lu-2015-a,Lu-2015-b} L\"u, Perkins, Pope, and Stelle identified
crucial sign errors which invalidate the non-trace part of Nelson's
proof. However, these authors still retained the validity of the trace
part, i.e. $\mathcal{R}\equiv0$ identically, which has tentatively
become known as the generalized Lichnerowicz theorem. \emph{If} this
``no-go''-type theorem is valid, then the constraint $\mathcal{R}=0$
that follows would nullify the contributions that stem from the $\mathcal{R}^{2}$
term to the quadratic-gravity field equation. Thus, for the \emph{sole}
purpose of finding static vacuo configurations, the $\mathcal{R}^{2}$
term may be suppressed. With $\beta$ set equal to zero and action
\eqref{eq:quadratic-action} reduced to an Einstein-Weyl gravity,
L\"u \emph{et al} then proceeded to discovering the L\"u-Perkins-Pope-Stelle
numerical solution, which represents a second branch of static, spherically
symmetric, and asymptotically flat spacetimes over and above the Schwarzschild
branch. This rather surprising result has generated considerable interest
\citep{Pravda-2018,Bonanno-2019,Goldstein-2017,Kokkotas-2017,Held-2022,Rinaldi-2020};
notably, Podolsk\'y \emph{et al} have identified an exact infinite-series
solution in place of the numerical approach \citep{Pravda-2018}.\vskip4pt

The generalized Lichnerowicz theorem was ``proved'' again by Kehagias
\emph{et al} \citep{Luest-2015-backholes} in a more limited situation,
the pure quadratic action, viz. $\gamma=0$. These authors concluded
that $\mathcal{R}\equiv\text{const}$ everywhere for this setup.\vskip4pt

However, these conclusions, put forth by Nelson \citep{Nelson-2010}
and re-enforced by L\"u \emph{et al} \citep{Lu-2015-a,Lu-2015-b}
and by Kehagias \emph{et al} \citep{Luest-2015-backholes},\linebreak
are under serious challenge. Inspired by a seminal -- yet obscure
-- sixty-year-old work by Buchdahl \citep{Buchdahl-1962}, we\linebreak
recently uncovered a new class of metrics that project \emph{non-constant}
scalar curvature in pure $\mathcal{R}^{2}$ gravity; the derivation
was detailed in our companion paper \citep{Nguyen-2022-Buchdahl}
and the result shall be summarized momentarily below. Leaving no stone
unturned, we successfully checked these Buchdahl-inspired metrics
directly against the pure $\mathcal{R}^{2}$ field equation, thereby
affirming their validity. The existence of the Buchdahl-inspired metrics
is in stark defiance of the generalized Lichnerowicz theorem.\vskip4pt

\emph{What has gone astray with the ``proofs'' of the generalized
Lichnerowicz theorem, which at first sight seem to be water-tight,
then?} It turns out that the ``proofs'' advocated in \citep{Nelson-2010,Lu-2015-a,Lu-2015-b,Luest-2015-backholes}
contain detrimental gaps which render the theorem vulnerable for violations.
Let us first expose the gaps in these ``proofs''.

\subsection*{The gaps in the generalized Lichnerowicz theorem}

We shall adopt the derivation of L\"u \emph{et al} in \citep{Lu-2015-a,Lu-2015-b}.
With $\alpha$, $\beta$, $\gamma$ all non-negative, the quadratic
action \eqref{eq:quadratic-action} produces the vacuo field equation
\small
\begin{align}
 & \gamma\Bigl(\mathcal{R}_{\mu\nu}-\frac{1}{2}g_{\mu\nu}\mathcal{R}\Bigr)-4\alpha B_{\mu\nu}\nonumber \\
 & \ \ \ +2\beta\biggl[\mathcal{R}\Bigl(\mathcal{R}_{\mu\nu}-\frac{1}{4}g_{\mu\nu}\mathcal{R}\Bigr)+\Bigl(g_{\mu\nu}\square-\nabla_{\mu}\nabla_{\nu}\Bigr)\mathcal{R}\biggr]=0\label{eq:field-eqn}
\end{align}
\normalsize where the Bach tensor $B_{\mu\nu}:=\left(\nabla^{\rho}\nabla^{\sigma}+\frac{1}{2}\mathcal{R}^{\rho\sigma}\right)\mathcal{C}_{\mu\nu\rho\sigma}$
is trace-free. The trace of \eqref{eq:field-eqn} yields:
\begin{equation}
-\gamma\,\mathcal{R}+6\beta\,\square\,\mathcal{R}=0\label{eq:trace-eqn}
\end{equation}

L\"u \emph{et al} then considered a static black hole metric of the
form $ds^{2}=-\lambda^{2}dt^{2}+h_{ij}dx^{i}dx^{j}$ where $\lambda$
and $h_{ij}$ are functions of the spatial coordinates, and $\lambda\geqslant0$
is supposed to vanish on the horizon. With the aid of \eqref{eq:trace-eqn},
for $\beta>0$, one obtains the following identity:\vskip-10pt

\small
\begin{equation}
\int_{V}d^{3}x\sqrt{h}D^{i}\left(\lambda\mathcal{R}D_{i}\mathcal{R}\right)=\int_{V}d^{3}x\sqrt{h}\lambda\Bigl[\left(D_{i}\mathcal{R}\right)^{2}+\frac{\gamma}{6\beta}\mathcal{R}^{2}\Bigr]\label{eq:identity}
\end{equation}
\normalsize 

\emph{If the left-hand-side of \eqref{eq:identity} could be made
vanish}, then the non-negativity of the right-hand-side of \eqref{eq:identity}
would readily enforce that
\begin{equation}
\mathcal{R}(r)\equiv\begin{cases}
\ 0 & \text{if }\gamma>0\\
\ \text{const} & \text{if }\gamma=0
\end{cases}\label{eq:Lichnerowicz}
\end{equation}
This constraint on static vacuo configurations for action \emph{\eqref{eq:quadratic-action}}
has tentatively become known as the generalized Lichnerowicz theorem.
Furthermore, restricted to the case of $\alpha=0$, the field equation
\eqref{eq:field-eqn}, subject to constraint \eqref{eq:Lichnerowicz},
leads to
\begin{equation}
\mathcal{R}_{\mu\nu}=\begin{cases}
\ 0 & \text{if }\gamma>0\\
\ \frac{1}{4}g_{\mu\nu}\times\text{const} & \text{if }\gamma=0
\end{cases}\label{eq:Lichnerowicz-2}
\end{equation}
meaning that the only static spherically symmetric vacuo admissible
for the $\mathcal{R}+\mathcal{R}^{2}$ action is Schwarzschild, and
likewise, the only static spherically symmetric vacuo admissible for
the pure $\mathcal{R}^{2}$ action is Schwarzschild-de Sitter. \vskip4pt

To make the left-hand-side of \eqref{eq:identity} vanish, in \citep{Nelson-2010},
with the integrand therein being a total derivative, Nelson applied
the 3D divergence theorem to turn the integral $\int_{V}d^{3}x\sqrt{h}D^{i}\left(\lambda\mathcal{R}D_{i}\mathcal{R}\right)$
into a surface term at spatial infinity. Next, he assumed that the
derivatives $D_{i}\mathcal{R}$ go to zero sufficiently rapidly so
that the surface term would vanish.\vskip4pt

Nelson's reasoning contains \emph{two} gaps, however. Firstly, an
actual vacuo configuration may \emph{not} guarantee that $D_{i}\mathcal{R}$
decay sufficiently rapidly to make the surface term vanish. This excessively
strong requirement can be susceptible for violations. Secondly, and
more seriously as this point is often overlooked, to apply the divergence
theorem, the integrand $D^{i}\left(\lambda\mathcal{R}D_{i}\mathcal{R}\right)$
must be a continuous function \emph{everywhere} within the integration
volume $V$. If the integrand diverges anywhere inside $V$, the divergence
theorem would cease to hold, and Nelson's proof would fall apart.
There is ample evidence that the second gap is pertinent: close to
a black hole, spacetimes often project singularities at the origin
and/or at the horizon.\vskip4pt

In \citep{Lu-2015-a,Lu-2015-b} L\"u \emph{et al} bypassed the second
gap in Nelson's proof by restricting the integration volume to the
\emph{exterior} of the black hole, viz. from the horizon outward to
the spatial infinity. In the exterior region, it is reasonable to
expect the integrand $D^{i}\left(\lambda\mathcal{R}D_{i}\mathcal{R}\right)$
to be well-behaved. As such, the left-hand-side of \eqref{eq:identity}
becomes the difference between \emph{two} surface terms, one at infinity
(denoted by $S_{\infty}$) and one at the horizon (denoted by $S_{h}$),
namely,
\begin{equation}
\oint_{S_{\infty}}d^{i}S\left(\lambda\mathcal{R}D_{i}\mathcal{R}\right)-\oint_{S_{h}}d^{i}S\left(\lambda\mathcal{R}D_{i}\mathcal{R}\right)\label{eq:diff-surface}
\end{equation}
L\"u \emph{et al}'s maneuver would seem to rescue Nelson's proof
from the abyss. However, their walk-around introduces a \emph{third}
gap: whereas $\lambda\rightarrow0$ on the horizon, the terms $\mathcal{R}D_{i}\mathcal{R}$
may diverge there and overwhelm $\lambda$, forcing the surface term
at the horizon $S_{h}$ to be finite or even divergent.\vskip4pt

In sum, due to the first gap and the third gap, the left-hand-side
of Eq. \eqref{eq:identity} in principle may -- and in practice can
-- deviate from zero, rendering the generalized Lichnerowicz theorem
invalid. \vskip4pt

As we shall see right below, there exists a class of non-Schwarzschild
vacua, the Buchdahl-inspired vacua, that project \emph{non-constant}
scalar curvature in stark violation of the \emph{would-be} conclusion
in Eq. \eqref{eq:Lichnerowicz} of the generalized Lichnerowicz theorem.

\section{\label{sec:Buchdahl-metric}The Buchdahl-inspired metric}

Circa 1962 Hans A. Buchdahl spearheaded a program seeking vacuum configurations
for pure $\mathcal{R}^{2}$ gravity, namely, the action \eqref{eq:quadratic-action}
with $\alpha=\gamma=0$ \citep{Buchdahl-1962}. Therein, he was able
to show that the pure $\mathcal{R}^{2}$ vacua in general can acquire
non-constant scalar curvature. Proceeding further, he arrived at a
non-linear second-order ordinary differential equation (ODE) which
would prescribe \emph{all} non-trivial static spherically symmetric
solutions admissible in the pure $\mathcal{R}^{2}$ theory. Unfortunately,
Buchdahl deemed his ODE intractable and prematurely discontinued his
pursuit for an analytical solution. To this day, his ODE remains untackled
and his 1962 work has largely gone unnoticed by the gravitation research
community \footnote{Buchdahl's paper has gathered a paltry sum of $40+$ citations since
its inception in 1962, according to the NASA ADS and InpireHEP databases.
Yet, none of these citations attempted to solve Buchdahl's ODE.}. Recently, in \citep{Nguyen-2022-Buchdahl} we revisited Buchdahl's
program, broke this outstanding six-decades-old impasse, and uncovered
a novel exhaustive class of vacua for pure $\mathcal{R}^{2}$ gravity,
to be summarized below.\vskip4pt

The Buchdahl-inspired metric, as we called it as such, takes the following
compact expression \vskip8pt\small

\noindent 
\begin{equation}
ds^{2}=e^{k\int\frac{dr}{r\,q(r)}}\left\{ p(r)\Bigl[-\frac{q(r)}{r}dt^{2}+\frac{r}{q(r)}dr^{2}\Bigr]+r^{2}d\Omega^{2}\right\} \label{eq:B-metric-1}
\end{equation}
\normalsize in which the pair of functions $\{p(r),q(r)\}$ obey
the ``evolution'' rules \small
\begin{align}
{\displaystyle \frac{dp}{dr}}\, & ={\displaystyle \,\frac{3k^{2}}{4\,r}\frac{p}{q^{2}}}\label{eq:B-metric-2}\\
{\displaystyle {\displaystyle \frac{dq}{dr}}}\, & =\,{\displaystyle \Bigl(1-\Lambda\,r^{2}\Bigr)\,p}\label{eq:B-metric-3}
\end{align}
\normalsize with the Ricci scalar equal
\begin{equation}
\mathcal{R}(r)\,=\,4\Lambda\,e^{-k\int\frac{dr}{r\,q(r)}}\label{eq:B-metric-4}
\end{equation}
The most crucial element of the metric is the new (Buchdahl) parameter
$k$ which allows the metric to be non-Schwarzschild and enables the
Ricci scalar to \emph{vary} on the manifold. At largest distances,
the Ricci scalar converges to $4\Lambda$. Metric \eqref{eq:B-metric-1}--\eqref{eq:B-metric-4}
is a bona fide enlargement of the Schwarzschild--de Sitter (SdS)
metric and duly recovers the SdS metric when $k=0$ (see the subsection
``The small $k$ limit'' right below). Our investigation on the
phase-space $\{p(r),q(r)\}$ of the evolution rules \eqref{eq:B-metric-2}--\eqref{eq:B-metric-3}
points towards very interesting new physics, e.g. the existence of
horizonless objects; our findings shall be presented in a separate
report.\vskip4pt

To alley any lingering doubt, in \citep{Nguyen-2022-Buchdahl} and
\citep{Shurtleff-2022}, the present author and Shurtleff successfully
checked the solution \eqref{eq:B-metric-1}--\eqref{eq:B-metric-4}
against the pure $\mathcal{R}^{2}$ vacuo field equation, thereby
affirming its validity. The existence of vacua with \emph{non-constant}
scalar curvature in a quadratic theory of gravity is a direct counterexample
against the generalized Lichnerowicz theorem stated in Eq. \eqref{eq:Lichnerowicz}.\vskip4pt

Moreover, to our surprise, despite being non-linear, the evolution
rules \eqref{eq:B-metric-2}--\eqref{eq:B-metric-3} are \emph{fully
soluble} for $\Lambda=0$. In a companion paper of this ``Beyond
Schwarzschild--de Sitter spacetimes'' series \citep{Nguyen-2022-Lambda0},
we exploited this advantage to derive an \emph{exact closed analytical
}form for a new metric, which we called the \emph{special} Buchdahl-inspired
metric that describes an asymptotically flat non-Schwarzschild $\mathcal{R}^{2}$
spacetime. The Kretschmann invariant of this metric exhibits curvature
singularities on the interior-exterior boundary. Novel anomalous behaviors
in the interior-exterior boundary and in the Kruskal-Szekeres diagram
of pure $\mathcal{R}^{2}$ spacetime structures are discovered and
reported in our companion paper \citep{Nguyen-2022-Lambda0}.

\subsection*{The small $k$ limit}

For a non-zero $\Lambda$ but with a small value of $k$, the Buchdahl-inspired
metric \eqref{eq:B-metric-1}--\eqref{eq:B-metric-4} admits a perturbative
form which we derived in \citep{Nguyen-2022-Buchdahl} and shall briefly
reproduce here for the reader's convenience. The crux of the argument
is that since the evolution rules \eqref{eq:B-metric-2}--\eqref{eq:B-metric-3}
depend on $k^{2}$ instead of $k$, they admit the perturbative solution
\begin{align}
p(r) & =1+\mathcal{O}\left(k^{2}\right)\label{eq:p-sol}\\
q(r) & =r-r_{\text{s}}-\frac{\Lambda}{3}r^{3}+\mathcal{O}\left(k^{2}\right)\label{eq:q-sol}
\end{align}
with $r_{\text{s}}$ being a constant. Note that the conformal factor
$e^{k\int\frac{dr}{r\,q(r)}}$ depends directly on $k$, however.
Plugging \eqref{eq:p-sol}--\eqref{eq:q-sol} into \eqref{eq:B-metric-1}
we obtain\vskip-10pt

\small
\begin{align}
ds^{2} & =e^{k\int\frac{dr}{r^{2}\Psi(r)}}\Bigl[-\Psi(r)dt^{2}+\frac{dr^{2}}{\Psi(r)}+r^{2}d\Omega^{2}\Bigr]+\mathcal{O}\left(k^{2}\right)\label{eq:small-k-B-metric}
\end{align}
\normalsize with $\Psi(r):=1-\frac{r_{\text{s}}}{r}-\frac{\Lambda}{3}r^{2}$
and the Ricci scalar given by
\begin{equation}
\mathcal{R}(r)=4\Lambda\,e^{-k\int\frac{dr}{r^{2}\,\Psi(r)}}+\mathcal{O}\left(k^{2}\right)\label{eq:-small-k-Ricci}
\end{equation}
Metric \eqref{eq:small-k-B-metric}--\eqref{eq:-small-k-Ricci} is
applicable for pure $\mathcal{R}^{2}$ gravity up to $\mathcal{O}\left(k^{2}\right)$,
with the Buchdahl parameter $k$ measuring the amount of deviation
from being Schwarzschild-de Sitter for the said metric. At $k=0$,
metric \eqref{eq:small-k-B-metric}--\eqref{eq:-small-k-Ricci} is
nothing but an SdS metric with a constant scalar curvature of $4\Lambda$.

\subsection*{The purpose of this paper}

Hereafter we shall concern with the following quadratic action
\begin{equation}
\mathcal{R}^{2}+\gamma\left(\mathcal{R}-2\Lambda\right)\label{eq:R-R2-L-action}
\end{equation}
We aim to show that the perturbative metric specified in Eq. \eqref{eq:small-k-B-metric}
for pure $\mathcal{R}^{2}$ gravity is \emph{extendible} to action
\eqref{eq:R-R2-L-action} upon a \emph{minor} modification. Inspired
by the expression \eqref{eq:small-k-B-metric}, we shall seek a metric
in the following form \vskip-10pt

\small
\begin{align}
ds^{2} & =e^{k\,\varphi(r)}\left[-\Psi(r)dt^{2}+\frac{dr^{2}}{\Psi(r)}+r^{2}d\Omega^{2}\right]+\mathcal{O}\left(k^{2}\right)\label{eq:a-3a}
\end{align}
\normalsize with $\Psi(r)$ still given by $1-\frac{r_{\text{s}}}{r}-\frac{\Lambda}{3}r^{2}$,
while $\varphi(r)$ \emph{is to be determined}. The case with $\varphi(r)\equiv0$
is obviously the classic SdS metric.\vskip8pt

The rest of our paper is organized as follows. In Sec. \ref{sec:Derivation}
we shall derive the perturbative vacuo \eqref{eq:a-3a} for action
\eqref{eq:R-R2-L-action} which is valid up to $\mathcal{O}\left(k^{2}\right)$,
with $k$ being a Buchdahl-like parameter reflecting the higher-derivative
nature of the action. Lemma \ref{lem:lemma-1} is the central result
of our paper. Next, we find the asymptotic limits for the new metric
at spatial infinity, in Sec. \ref{sec:Large-distance}. We then embed
the new metric in the larger context of full quadratic-gravity theory,
viz. action \eqref{eq:quadratic-action}, with emphasis on its connection
to the L\"u-Perkins-Pope-Stelle solution in Einstein-Weyl gravity,
in Sec. \ref{sec:Implications}.

\section{\label{sec:Derivation}Constructing a perturbative\vskip2pt vacuo
for the $\boldsymbol{R^{2}+R+\Lambda}$ action}

The quadratic action in \eqref{eq:R-R2-L-action} has the vacuo field
equation \small
\begin{align}
2\left[\mathcal{R}\Bigl(\mathcal{R}_{\mu\nu}-\frac{1}{4}g_{\mu\nu}\mathcal{R}\Bigr)+\Bigl(g_{\mu\nu}\square-\nabla_{\mu}\nabla_{\nu}\Bigr)\mathcal{R}\right]\ \ \ \nonumber \\
+\gamma\Bigl(\mathcal{R}_{\mu\nu}-\frac{1}{2}g_{\mu\nu}\mathcal{R}+\Lambda g_{\mu\nu}\Bigr) & =0\label{eq:a-3c}
\end{align}
\normalsize Upon taking the trace
\begin{equation}
6\,\square\,\mathcal{R}-\gamma\,\left(\mathcal{R}-4\Lambda\right)=0\label{eq:a-3d}
\end{equation}
to get rid of the cumbersome $\,\square\,\mathcal{R}$ term, Eq. \eqref{eq:a-3c}
is transformed into \small
\begin{align}
\mathcal{R}\Bigl(\mathcal{R}_{\mu\nu}-\frac{1}{4}g_{\mu\nu}\mathcal{R}\Bigr)-\nabla_{\mu}\nabla_{\nu}\mathcal{R}\ \ \ \ \ \ \ \ \ \ \nonumber \\
+\frac{\gamma}{2}\Bigl(\mathcal{R}_{\mu\nu}-\frac{1}{6}g_{\mu\nu}\mathcal{R}-\frac{\Lambda}{3}g_{\mu\nu}\Bigr) & =0\label{eq:a-3e}
\end{align}
\normalsize

With the metric components dependent on $r$, the field equation \eqref{eq:a-3e}
has three remaining independent components against one unknown function
$\varphi(r)$ (while $\Phi(r)$ has been fixed to be $1-\frac{r_{\text{s}}}{r}-\frac{\Lambda}{3}r^{2}$).
At this stage, the problem appears to be over-determined; this mirage
will resolve itself, as we shall see. The relevant components of the
fourth-order derivative term are \footnote{Recall that for a scalar field $\phi$: $\nabla_{\mu}\nabla_{\nu}\phi=\partial_{\mu}\partial_{\nu}\phi-\Gamma_{\mu\nu}^{\lambda}\partial_{\lambda}\phi$.}
\begin{align}
\nabla_{0}\nabla_{0}\mathcal{R} & =-\Gamma_{00}^{1}\mathcal{R}'(r)\\
\nabla_{1}\nabla_{1}\mathcal{R} & =-\Gamma_{11}^{1}\mathcal{R}'(r)+\mathcal{R}''(r)\\
\nabla_{2}\nabla_{2}\mathcal{R} & =-\Gamma_{22}^{1}\mathcal{R}'(r)
\end{align}
The $rr$-component involves the second-derivative of $\mathcal{R}$
with respect to $r$ and is thus quite cumbersome to deal with. Therefore,
in place of the $rr$-component of the field equation, we shall use
the trace equation as a surrogate for it, together with the $tt$-
and $\theta\theta$- components in our calculations below.\vskip4pt

The key result is summarized in Lemma \ref{lem:lemma-1}.\vskip4pt
\begin{lem}
\label{lem:lemma-1}The line element
\begin{equation}
ds^{2}=e^{k\,\varphi(r)}\biggl[-\Psi(r)dt^{2}+\frac{dr^{2}}{\Psi(r)}+r^{2}d\Omega^{2}\biggr]\label{eq:O-k-metric}
\end{equation}
with $\Psi(r):=1-\frac{r_{\text{s}}}{r}-\frac{\Lambda}{3}r^{2}$ and
$\varphi(r)$ obeying
\begin{equation}
6\left(r^{2}\Psi(r)\varphi'(r)\right)'=\gamma\,r^{2}\varphi(r)\label{eq:varphi-eqn}
\end{equation}
satisfies the field equation \eqref{eq:a-3c} up to $\mathcal{O}\left(k^{2}\right)$.
\end{lem}
\vskip12pt
\begin{proof}
We shall use Eq. \eqref{eq:a-3e} in place of Eq. \eqref{eq:a-3c}.
The $tt-$ and $\theta\theta-$ components of Eq. \eqref{eq:a-3e}
can be cast as \small
\begin{align}
 & \frac{\mathcal{R}_{00}}{\Psi}-\frac{1}{4}\frac{g_{00}e^{-\varphi}}{\Psi}\left(\mathcal{R}e^{\varphi}\right)+\frac{\Gamma_{00}^{1}}{\Psi}\frac{\mathcal{R}'}{\mathcal{R}}\nonumber \\
 & \ \ \ \ \ =-\frac{\gamma}{2\mathcal{R}}\left[\frac{\mathcal{R}_{00}}{\Psi}-\frac{1}{6}\frac{g_{00}e^{-\varphi}}{\Psi}\left(\mathcal{R}e^{\varphi}\right)-\frac{\Lambda}{3}\frac{g_{00}e^{-\varphi}}{\Psi}e^{\varphi}\right]\label{eq:a-4a}\\
 & \frac{\mathcal{R}_{22}}{r^{2}}-\frac{1}{4}\frac{g_{22}e^{-\varphi}}{r^{2}}\left(\mathcal{R}e^{\varphi}\right)+\frac{\Gamma_{22}^{1}}{r^{2}}\frac{\mathcal{R}'}{\mathcal{R}}\nonumber \\
 & \ \ \ \ \ =-\frac{\gamma}{2\mathcal{R}}\left[\frac{\mathcal{R}_{22}}{r^{2}}-\frac{1}{6}\frac{g_{22}e^{-\varphi}}{r^{2}}\left(\mathcal{R}e^{\varphi}\right)-\frac{\Lambda}{3}\frac{g_{22}e^{-\varphi}}{r^{2}}e^{\varphi}\right]\label{eq:a-4b}
\end{align}
\normalsize The trace equation \eqref{eq:a-3d} is recast as
\begin{equation}
6(e^{\varphi}r^{2}\Psi\mathcal{R}')'=\gamma\,r^{2}e^{2\varphi}\left(\mathcal{R}-4\Lambda\right)\label{eq:a-4c}
\end{equation}
In all calculations below, the $\approx$ sign when used means that
we keep only up to the first order terms in $k$. The relevant tensor
components are\small
\begin{align}
\frac{\mathcal{R}_{00}}{\Psi} & =\Bigl(\frac{k\varphi''}{2}+\frac{k^{2}\varphi'^{2}}{2}+\frac{k\varphi'}{r}\Bigr)\Psi+\frac{\Psi''}{2}+\Bigl(k\varphi'+\frac{1}{r}\Bigr)\Psi'\\
 & \approx\Bigl(\frac{\Psi''}{2}+\frac{\Psi'}{r}\Bigr)+k\,\Bigl(\frac{\varphi''\Psi}{2}+\frac{\varphi'\Psi}{r}+\varphi'\Psi'\Bigr)\\
 & =-\Lambda+k\,\Bigl(\frac{\varphi''\Psi}{2}+\frac{\varphi'\Psi}{r}+\varphi'\Psi'\Bigr)\label{eq:a-5a}
\end{align}
\begin{align}
\frac{\mathcal{R}_{22}}{r^{2}} & =\frac{1}{r^{2}}-\Bigl(\frac{k\varphi''}{2}+\frac{k^{2}\varphi'^{2}}{2}+\frac{2k\varphi'}{r}+\frac{1}{r^{2}}\Bigr)\Psi\nonumber \\
 & \ \ \ \ \ \ \ \ \ \ -\Bigl(\frac{k\varphi'}{2}+\frac{1}{r}\Bigr)\Psi'\\
 & \approx\Bigl(\frac{1}{r^{2}}-\frac{\Psi}{r^{2}}-\frac{\Psi'}{r}\Bigr)-k\,\Bigl(\frac{\varphi''\Psi}{2}+\frac{2\varphi'\Psi}{r}+\frac{\varphi'\Psi'}{2}\Bigr)\\
 & =\Lambda-k\,\Bigl(\frac{\varphi''\Psi}{2}+\frac{2\varphi'\Psi}{r}+\frac{\varphi'\Psi'}{2}\Bigr)\label{eq:a-5b}
\end{align}
\begin{align}
\mathcal{R}e^{k\,\varphi} & =\frac{2}{r^{2}}-\Bigl(3k\varphi''+\frac{3k^{2}\varphi'^{2}}{2}+\frac{6k\varphi'}{r}+\frac{2}{r^{2}}\Bigr)\Psi\nonumber \\
 & \ \ \ \ \ \ \ \ \ \ \ \ \ \ \ \ \ \ \ \ -\Bigl(3k\varphi'+\frac{4}{r}\Bigr)\Psi'-\Psi''\\
 & \approx\Bigl(\frac{2}{r^{2}}-\frac{2\Psi}{r^{2}}-\frac{4\Psi'}{r}-\Psi''\Bigr)\nonumber \\
 & \ \ \ \ \ \ \ \ \ \ \ \ \ \ \ -k\,\Bigl(3\varphi''\Psi+\frac{6\varphi'\Psi}{r}+3\varphi'\Psi'\Bigr)\\
 & =4\Lambda-k\,\Bigl(3\varphi''\Psi+\frac{6\varphi'\Psi}{r}+3\varphi'\Psi'\Bigr)\label{eq:a-5c}
\end{align}
\begin{align}
\mathcal{R} & =(\mathcal{R}e^{k\,\varphi})e^{-k\,\varphi}\\
 & \approx4\Lambda-4\Lambda k\varphi-k\,\Bigl(3\varphi''\Psi+\frac{6\varphi'\Psi}{r}+3\varphi'\Psi'\Bigr)\label{eq:a-5d}
\end{align}
\begin{align}
\mathcal{R}'e^{k\,\varphi} & =\left(\mathcal{R}e^{k\,\varphi}\right)'-k\varphi'\left(\mathcal{R}e^{k\,\varphi}\right)\\
 & \approx-k\,\Bigl(3\varphi''\Psi+\frac{6\varphi'\Psi}{r}+3\varphi'\Psi'\Bigr)'-4\Lambda k\varphi'\label{eq:a-5e}\\
\frac{\mathcal{R}'}{\mathcal{R}} & \approx-k\varphi'-\frac{k}{4\Lambda}\Bigl(3\varphi''\Psi+\frac{6\varphi'\Psi}{r}+3\varphi'\Psi'\Bigr)'\label{eq:a-5f}
\end{align}
and
\begin{align}
\frac{g_{00}e^{-k\,\varphi}}{\Psi} & =-1\\
\frac{g_{22}e^{-k\,\varphi}}{r^{2}} & =1\\
\frac{\Gamma_{00}^{1}}{\Psi}\  & =k\frac{\varphi'\Psi}{2}+\frac{\Psi'}{2}\\
\frac{\Gamma_{22}^{1}}{r^{2}}\  & =-\left(k\frac{\varphi'}{2}+\frac{1}{r}\right)\Psi
\end{align}
\normalsize The left-hand-side of Eqs. \eqref{eq:a-4a} and \eqref{eq:a-4b}
are\small
\begin{align}
 & \frac{\mathcal{R}_{00}}{\Psi}-\frac{1}{4}\frac{g_{00}e^{-k\,\varphi}}{\Psi}\left(\mathcal{R}e^{k\,\varphi}\right)+\frac{\Gamma_{00}^{1}}{\Psi}\frac{\mathcal{R}'}{\mathcal{R}}\nonumber \\
 & \ \ \ \approx-\frac{k}{4}\left(\varphi''\Psi+2\frac{\varphi'\Psi}{r}+\varphi'\Psi'\right)\nonumber \\
 & \ \ \ \ \ \ \ \ -\frac{\Psi'}{8\Lambda}k\,\Bigl(3\varphi''\Psi+\frac{6\varphi'\Psi}{r}+3\varphi'\Psi'\Bigr)'\\
 & \ \ \ =-\frac{k}{4}\frac{\left(\varphi'r^{2}\Psi\right)'}{r^{2}}-\frac{3\Psi'}{8\Lambda}k\left(\frac{\left(\varphi'r^{2}\Psi\right)'}{r^{2}}\right)'\label{eq:a-6a}
\end{align}
\begin{align}
 & \frac{\mathcal{R}_{22}}{r^{2}}-\frac{1}{4}\frac{g_{22}e^{-k\,\varphi}}{r^{2}}\left(\mathcal{R}e^{k\,\varphi}\right)+\frac{\Gamma_{22}^{1}}{r^{2}}\frac{\mathcal{R}'}{\mathcal{R}}\nonumber \\
 & \ \ \ \approx\frac{k}{4}\left(\varphi''\Psi+2\frac{\varphi'\Psi}{r}+\varphi'\Psi'\right)\nonumber \\
 & \ \ \ \ \ \ \ \ +\frac{\Psi}{4\Lambda r}k\Bigl(3\varphi''\Psi+\frac{6\varphi'\Psi}{r}+3\varphi'\Psi'\Bigr)'\\
 & \ \ \ =\frac{k}{4}\frac{\left(\varphi'r^{2}\Psi\right)'}{r^{2}}+\frac{3\Psi}{4\Lambda r}k\left(\frac{\left(\varphi'r^{2}\Psi\right)'}{r^{2}}\right)'\label{eq:a-6b}
\end{align}
\normalsize The bracketed terms in the right-hand-side of Eqs. \eqref{eq:a-4a}
and \eqref{eq:a-4b} are\small
\begin{align}
\frac{\mathcal{R}_{00}}{\Psi}-\frac{1}{6}\frac{g_{00}e^{-k\,\varphi}}{\Psi}\left(\mathcal{R}e^{k\,\varphi}\right)-\frac{\Lambda}{3}\frac{g_{00}e^{-k\,\varphi}}{\Psi}e^{k\,\varphi}\ \ \ \nonumber \\
\approx k\frac{\varphi'\Psi'}{2}+\frac{\Lambda}{3}k\varphi\\
\frac{\mathcal{R}_{22}}{r^{2}}-\frac{1}{6}\frac{g_{22}e^{-k\,\varphi}}{r^{2}}\left(\mathcal{R}e^{k\,\varphi}\right)-\frac{\Lambda}{3}\frac{g_{22}e^{-k\,\varphi}}{r^{2}}e^{k\,\varphi}\ \ \ \nonumber \\
\approx-k\frac{\varphi'\Psi}{r}-\frac{\Lambda}{3}k\varphi
\end{align}
\normalsize The terms in the trace equation \eqref{eq:a-4c} are\small
\begin{align}
e^{k\,\varphi}r^{2}\Psi\mathcal{R}' & =(\mathcal{R}e^{k\,\varphi})r^{2}\Psi\frac{\mathcal{R}'}{\mathcal{R}}\\
 & \approx4\Lambda kr^{2}\Psi\left[-\varphi'-\frac{1}{4\Lambda}\Bigl(3\varphi''\Psi+\frac{6\varphi'\Psi}{r}+3\varphi'\Psi'\Bigr)'\right]\\
 & =-4\Lambda k\varphi'r^{2}\Psi-3kr^{2}\Psi\left(\frac{\left(\varphi'r^{2}\Psi\right)'}{r^{2}}\right)'
\end{align}
\begin{align}
r^{2}e^{2k\,\varphi}\left(\mathcal{R}-4\Lambda\right) & \approx kr^{2}\left[-4\Lambda\varphi-\Bigl(3\varphi''\Psi+\frac{6\varphi'\Psi}{r}+3\varphi'\Psi'\Bigr)\right]\\
 & =-4\Lambda kr^{2}\varphi-3k\left(\varphi'r^{2}\Psi\right)'
\end{align}
\normalsize Up to the first-order in $k$, Eqs. \eqref{eq:a-4a},
\eqref{eq:a-4b}, and \eqref{eq:a-4c} are\small
\begin{align}
\frac{1}{4}\frac{\left(\varphi'r^{2}\Psi\right)'}{r^{2}}+\frac{3\Psi'}{8\Lambda}\left(\frac{\left(\varphi'r^{2}\Psi\right)'}{r^{2}}\right)' & =\frac{\gamma}{8\Lambda}\left(\frac{\varphi'\Psi'}{2}+\frac{\Lambda}{3}\varphi\right)\\
\frac{1}{4}\frac{\left(\varphi'r^{2}\Psi\right)'}{r^{2}}+\frac{3\Psi}{4\Lambda r}\left(\frac{\left(\varphi'r^{2}\Psi\right)'}{r^{2}}\right)' & =\frac{\gamma}{8\Lambda}\left(\frac{\varphi'\Psi}{r}+\frac{\Lambda}{3}\varphi\right)
\end{align}
\vskip-12pt
\begin{align}
6\Biggl(4\Lambda\varphi'r^{2}\Psi3r^{2}\Psi\biggl(\frac{\bigl(\varphi'r^{2}\Psi\bigr)'}{r^{2}}\biggr)'\Biggr)' & =\gamma\left[4\Lambda r^{2}\varphi+3\bigl(\varphi'r^{2}\Psi\bigr)'\right]
\end{align}
\normalsize which, upon rearranging, become \small
\begin{align}
6\frac{(\varphi'r^{2}\Psi)'}{r^{2}}-\gamma\varphi & =-\frac{3\Psi'}{2\Lambda}\biggl(6\frac{(\varphi'r^{2}\Psi)'}{r^{2}}-\gamma\varphi\biggr)'\label{eq:final-1}\\
6\frac{(\varphi'r^{2}\Psi)'}{r^{2}}-\gamma\varphi & =-\frac{3\Psi}{\Lambda r}\biggl(6\frac{(\varphi'r^{2}\Psi)'}{r^{2}}-\gamma\varphi\biggr)'\label{eq:final-2}\\
6\frac{(\varphi'r^{2}\Psi)'}{r^{2}}-\gamma\varphi & =-\frac{3}{4\Lambda r^{2}}\Biggl(r^{2}\Psi\biggl(6\frac{(\varphi'r^{2}\Psi)'}{r^{2}}-\gamma\varphi\biggr)'\Biggr)'\label{eq:final-3}
\end{align}
\normalsize Remarkably, all three equations \eqref{eq:final-1}--\eqref{eq:final-3}
are automatically satisfied \emph{if and only if} $\varphi(r)$ obeys
the following linear second-order ODE
\begin{align}
6\left(r^{2}\Psi\varphi'\right)' & =\gamma\,r^{2}\varphi
\end{align}
Metric \eqref{eq:O-k-metric}--\eqref{eq:varphi-eqn} is thus established.
\end{proof}
\begin{rem}
Let us ignore the conformal factor in expression \eqref{eq:O-k-metric}
for the moment. The terms in the square bracket of \eqref{eq:O-k-metric}
is an SdS metric -- a ``seed'' metric -- with a constant scalar
curvature of $4\Lambda.$ The Ricci scalar of metric \eqref{eq:O-k-metric}
can be obtained via a conformal transformation from the ``seed''
SdS metric \citep{Shapiro-2004}, per
\begin{align}
\mathcal{R} & =e^{-k\,\varphi}\left(4\Lambda-3k\,\tilde{\square}\varphi\right)+\mathcal{O}\left(k^{2}\right)\label{eq:Ricci}
\end{align}
with the tilde denoting derivatives using the ``seed'' SdS metric.
By virtue of \eqref{eq:varphi-eqn}
\begin{align}
\tilde{\square}\varphi & =\frac{1}{\sqrt{-\tilde{g}}}\partial_{r}\left(\sqrt{-\tilde{g}}\tilde{g}^{rr}\partial_{r}\varphi\right)\\
 & =\frac{1}{r^{2}\sin\theta}\left(r^{2}\sin\theta\,\Psi\varphi'\right)'\\
 & =\frac{\gamma}{6}\varphi
\end{align}
The Ricci scalar \eqref{eq:Ricci} is thus
\begin{align}
\mathcal{R}(r) & =\left(1-k\,\varphi\right)\left(4\Lambda-\frac{\gamma}{2}k\,\varphi\right)+\mathcal{O}\left(k^{2}\right)\\
 & =4\Lambda-k\left(4\Lambda+\frac{\gamma}{2}\right)\varphi(r)+\mathcal{O}\left(k^{2}\right)\label{eq:Ricci-result}
\end{align}
\end{rem}
\vskip6pt
\begin{rem}
Using the symbolic manipulator MAXIMA ONLINE, we were able to verify
the result in \eqref{eq:Ricci-result} concerning the Ricci scalar.
We also verified that the trace equation \eqref{eq:a-3d} and the
$tt-$, $rr-$, $\theta\theta-$ components of the field equation
\eqref{eq:a-3e} vanish \emph{up to} $\mathcal{O}\left(k^{2}\right)$.
\end{rem}
\vskip6pt
\begin{rem}
\label{rem:non-const-Ricci}Per \eqref{eq:Ricci-result}, the Ricci
scalar is \emph{non-constant}, \emph{including} the case with $\Lambda=0$,
as long as $k\neq0$ and $\gamma\neq0$.\linebreak The parameter
$k$ therefore acts as an equivalent to the Buchdahl parameter used
in metric \eqref{eq:B-metric-1}--\eqref{eq:B-metric-4} and metric
\eqref{eq:small-k-B-metric}--\eqref{eq:-small-k-Ricci} in the pure
$\mathcal{R}^{2}$ action.
\end{rem}
\vskip6pt
\begin{rem}
\noindent The case of $\gamma=0$ for action \eqref{eq:R-R2-L-action}
amounts to pure $\mathcal{R}^{2}$ gravity. From \eqref{eq:varphi-eqn},
we have $r^{2}\Psi\varphi'=\text{const}$, yielding $\varphi=\int\frac{dr}{r^{2}\Psi}$
in perfect agreement with the small-$k$ expansion of the Buchdahl-inspired
metric, per Eq. \eqref{eq:small-k-B-metric}. The case of $\gamma=\infty$
for action \eqref{eq:R-R2-L-action} amounts to the Einstein-Hilbert
action augmented with a cosmological constant. Eq. \eqref{eq:varphi-eqn}
then enforces that $\varphi=0$, thence reproducing the SdS metric
as expected.
\end{rem}
\vskip6pt
\begin{rem}
The ODE \eqref{eq:varphi-eqn} is of second differential order. It
entails two boundary conditions. For the case of $\Lambda\leqslant0$,
in which the radial coordinate $r$ is in the range $(0,\infty)$,
we are at liberty to set $\varphi(r\rightarrow\infty)=0$. Since the
ODE is linear, the magnitude of $\varphi$ can be absorbed into the
Buchdahl parameter $k$ in the conformal factor of Eq. \eqref{eq:O-k-metric}.
\end{rem}

\section{\label{sec:Large-distance}Large-distance asymptotic\vskip2pt behavior}

\subsubsection{\textbf{The case of $\Lambda<0$}}

\noindent As $r\rightarrow\infty$, Eq. \eqref{eq:varphi-eqn} asymptotically
is
\begin{align}
2|\Lambda|\left(r^{4}\varphi'\right)' & \simeq\gamma\,r^{2}\varphi
\end{align}
which is soluble. Substituting $x=\ln r$, the equation above becomes
\begin{equation}
\frac{d^{2}\varphi}{dx^{2}}+3\frac{d\varphi}{dx}-\frac{\gamma}{2|\Lambda|}\varphi=0
\end{equation}
which has the solution:
\begin{align}
\varphi(x) & =\xi_{+}e^{\left(-\frac{3}{2}+\sqrt{\frac{9}{4}+\frac{\gamma}{2|\Lambda|}}\right)x}+\xi_{-}e^{\left(-\frac{3}{2}-\sqrt{\frac{9}{4}+\frac{\gamma}{2|\Lambda|}}\right)x}\label{eq:a-asymp-1}
\end{align}

\noindent Since $\gamma>0$, the first term in \eqref{eq:a-asymp-1}
has a growing exponent and should be discarded. The asymptotic is
left with
\begin{equation}
\varphi(r)\simeq r^{-\frac{3}{2}\left(1+\sqrt{1+\frac{2\gamma}{9|\Lambda|}}\right)}
\end{equation}
which vanishes at spatial infinity. The metric is asymptotically anti-de
Sitter.

\subsubsection{\textbf{\label{subsec:Asymp-flat}The case of $\Lambda=0$}}

\noindent As $r\rightarrow\infty$, Eq. \eqref{eq:varphi-eqn} asymptotically
is
\begin{align}
6\left(r^{2}\varphi'\right)' & \simeq\gamma\,r^{2}\varphi
\end{align}
Using $\varphi(r):=\chi(r)/r$, the equation for $\chi(r)$ is soluble,
giving the asymptotic
\begin{align}
\varphi(x) & \simeq\frac{e^{-\sqrt{\frac{\gamma}{6}}r}}{r}
\end{align}
which vanishes at spatial infinity. The metric is \emph{asymptotically
flat}.\vskip4pt

This case is particularly interesting. The metric is asymptotically
flat; yet, \emph{in the bulk}, it develops non-constant scalar curvature.
The generalized Lichnerowicz theorem is evaded in its entirety.

\section{\label{sec:Implications}Implications for the\vskip1pt L\"u-Perkins-Pope-Stelle
solution}

In \citep{Lu-2015-a,Lu-2015-b}, influenced by the generalized Lichnerowicz
theorem, L\"u, Perkins, Pope and Stelle suppressed the $\mathcal{R}^{2}$
term in their exploration of black hole configurations for quadratic
gravity, viz. action \eqref{eq:quadratic-action}. Their reason was
that, \emph{provided that }the generalized Lichnerowicz theorem were
valid, a vanishing $\mathcal{R}$ would automatically kill off the
terms in the square bracket -- which are associated with $\beta$
-- in the second line of the field equation \eqref{eq:field-eqn}.
The terms in question are the contributions of the $\mathcal{R}^{2}$
term of action \eqref{eq:quadratic-action} to the field equation.
Accordingly, \emph{solely} for the purpose of finding \emph{static}
vacua, in \emph{assuming} the validity of the generalized Lichnerowicz
theorem, it would have been legitimate to set $\beta$ equal to zero
\footnote{Note that for \emph{other} purposes, such as finding \emph{non-static}
vacua, $\beta$ must be restored into the investigation, however}. This was indeed what L\"u \emph{et al} did. They went on to discover
the L\"u-Perkins-Pope-Stelle numerical solution for the leftover
Einstein-Weyl gravity. In \citep{Pravda-2018} Podolsk\'y \emph{et
al} followed up with an exact infinite-series solution in place of
the numerical solution.\vskip4pt

However, as established in our previous work \citep{Nguyen-2022-Buchdahl}
and in Lemma \ref{lem:lemma-1} herein, the existence of the class
of Buchdahl-inspired metrics in pure $\mathcal{R}^{2}$ gravity alongside
with the perturbative vacuo in the $\mathcal{R}^{2}+\gamma(\mathcal{R}-2\Lambda)$
action \emph{denies\linebreak} the generalized Lichnerowicz theorem
in \emph{its entirety}. These spacetimes project \emph{non-constant}
scalar curvature. The $\mathcal{R}^{2}$ term must be restored into
action \eqref{eq:quadratic-action}, viz. $\gamma\,\mathcal{R}+\beta\,\mathcal{R}^{2}-\alpha\,\mathcal{C}^{\mu\nu\rho\sigma}\mathcal{C}_{\mu\nu\rho\sigma}$,
for the purpose of finding static vacuo configurations. That is to
say, unless one can suppress $\beta$ by some other theoretical or
observational reason, the generalized Lichnerowicz theorem -- having
lost its legitimacy -- is \emph{not} a justifiable cause to kill
off $\beta$.\vskip4pt

With $\beta$ being reinstated, an immediate consequence would be
to extend the L\"u-Perkins-Pope-Stelle ansatz in \citep{Lu-2015-a,Lu-2015-b}
to the \emph{full} quadratic action. Equivalently, the infinite-series
approach pursued by Podolsk\'y \emph{et al} in \citep{Pravda-2018}
could be suitable for an extension with $\beta\neq0$.\vskip4pt

A more modest setup would be to rework the L\"u-Perkins-Pope-Stelle
ansatz (or that of Podolsk\'y \emph{et al}) for the $\gamma\,\mathcal{R}+\beta\,\mathcal{R}^{2}$
action, i.e., by excluding the Weyl term. This theory is ghost-free
and is equivalent to the standard Einstein gravity with one additional
scalar degree of freedom \citep{Stelle-1978,Stelle-1977}. Regarding
the L\"u-Perkins-Pope-Stelle ansatz, for the $\gamma\,\mathcal{R}+\beta\,\mathcal{R}^{2}$
action, the mass $m_{0}:=\sqrt{\gamma/(6\beta)}$ of the massive spin-0
mode would stand in place for the mass $m_{2}:=\sqrt{\gamma/(2\alpha)}$
of the massive spin-2 mode which is now absent. Advantages in exploring
the $\mathcal{R}+\mathcal{R}^{2}$ action would be to deal with a
considerably simpler field equation, and that the issues with ghosts
would stay silent. \vskip4pt

Despite the absence of the Bach tensor in its field equation, the
$\gamma\,\mathcal{R}+\beta\,\mathcal{R}^{2}$ action should already
project very rich phenomenology. The reason is that the vacua of this
theory should inherit some properties of the L\"u-Perkins-Pope-Stelle
solution \emph{and} those of the Buchdahl-inspired solution, the latter
being able to participate owing to the $\beta\,\mathcal{R}^{2}$ component
in the action. Note that the two said solutions are of complementary
nature. Whereas the L\"u-Perkins-Pope-Stelle solution represents
a second branch of static, spherically symmetric, and asymptotically
flat spacetimes \emph{separate} from the Schwarzschild branch, the
Buchdahl-inspired solution \emph{supersedes} the Schwarzschild branch.\vskip4pt

For the $\gamma\,\mathcal{R}+\beta\,\mathcal{R}^{2}$ action, the
$\mathcal{O}\left(k^{2}\right)$ perturbative result obtained in Lemma
\ref{lem:lemma-1} would already provide a useful guidepost. The full
solution -- yet to be determined -- needs to recover metric \eqref{eq:O-k-metric}--\eqref{eq:varphi-eqn}
in the limit of small $k$. An important question to find out is how
the Buchdahl parameter $k$ in \eqref{eq:O-k-metric}--\eqref{eq:varphi-eqn}
is translated into the built-in degree of ``non-Schwarzschildness''
in the L\"u-Perkins-Pope-Stelle ansatz.\vskip4pt

There is one serious caveat. As we briefly alluded to in Sec. \ref{sec:Buchdahl-metric},
in the asymptotic flatness limit, the Buchdahl-inspired metric given
by Eqs. \eqref{eq:B-metric-1}--\eqref{eq:B-metric-4} admits an
\emph{exact closed analytical} form, which we called the \emph{special}
Buchdahl-inspired metric. Our detailed derivation is presented in
our companion paper \citep{Nguyen-2022-Lambda0}. For the reader's
convenience, we reproduce the \emph{special }Buchdahl-inspired metric
below:

\footnotesize
\begin{equation}
ds^{2}=\left|1-\frac{r_{\text{s}}}{\rho}\right|^{\frac{k}{r_{\text{s}}}}\biggl\{-\Bigl(1-\frac{r_{\text{s}}}{\rho}\Bigr)dt^{2}+\frac{r^{4}(\rho)\,d\rho^{2}}{\rho^{4}\Bigl(1-\frac{r_{\text{s}}}{\rho}\Bigr)}+r^{2}(\rho)\,d\Omega^{2}\biggr\}\label{eq:special-B-1}
\end{equation}
\normalsize in which $\rho$ is the radial coordinate and the \emph{areal}
coordinate $r$ is given by \footnotesize
\begin{equation}
r(\rho)=\frac{\zeta\,r_{\text{s}}\left|1-\frac{r_{\text{s}}}{\rho}\right|^{\frac{1}{2}\left(\zeta-1\right)}}{1-\text{sgn}\Bigl(1-\frac{r_{\text{s}}}{\rho}\Bigr)\left|1-\frac{r_{\text{s}}}{\rho}\right|^{\zeta}};\ \ \ \zeta:=\sqrt{1+3k^{2}/r_{\text{s}}^{2}}\label{eq:special-B-2}
\end{equation}
\normalsize It describes a static spherically symmetric $\mathcal{R}^{2}$
structure that lives on an asymptotically flat spacetime. The structure
is found to possess\emph{ singular} (i.e., non-analytic and anomalous)
behaviors across its interior-exterior boundary as well as in its
Kruskal-Szekeres diagram, with the Buchdahl parameter $k$ being the
root cause of all these anomalies; see our companion paper for a detailed
exposition \citep{Nguyen-2022-Lambda0}.\vskip4pt

Back to the $\mathcal{R}+\mathcal{R}^{2}$ action at hand. The full
vacuo solution to the $\gamma\,\mathcal{R}+\beta\,\mathcal{R}^{2}$
action must approach the \emph{special} Buchdahl-inspired metric \eqref{eq:special-B-1}--\eqref{eq:special-B-2}
in the limit of $\gamma=0$ and $\Lambda=0$ (i.e., asymptotic flatness).
The L\"u-Perkins-Pope-Stelle ansatz \emph{in its current form} lacks
a degree of non-analyticity necessary to recover the \emph{special}
Buchdahl-inspired metric. In order to succeed, the L\"u-Perkins-Pope-Stelle
ansatz would need to have some non-analytic built-in ingredients to
be able to accommodate the powers of $\left|1-\frac{r_{\text{s}}}{\rho}\right|$
in Eqs. \eqref{eq:special-B-1} and \eqref{eq:special-B-2}. \vskip4pt

Put another way, on the one hand, in the regime of $m_{0}\rightarrow0$
(i.e., $\gamma\rightarrow0$ while $\beta$ is fixed), the singular
footprints of the \emph{special} Buchdahl-inspired metric by way of
$\left|1-\frac{r_{\text{s}}}{\rho}\right|^{\zeta}$ and similar terms
in \eqref{eq:special-B-1}--\eqref{eq:special-B-2} should manifest
in the full solution -- yet to be identified -- for regions close
to the interior-exterior boundary. Note that the auxiliary parameter
$\zeta$ is not necessarily a rational number. On the other hand,
in the regime of $m_{0}\rightarrow\infty$ (i.e. $\beta\rightarrow0$
while $\gamma$ stays put), the classic Schwarzschild solution should
become dominant in the full solution.\vskip4pt

A tantalizing question arises: \emph{What is the nature of the transition
point between the two regimes, as $m_{0}$ is tuned from zero to infinity?}

\section{\label{sec:Conclusions}Conclusions}

Lemma \ref{lem:lemma-1} in Sec. \ref{sec:Derivation} is the central
result of our paper. Inspired by the Buchdahl-inspired metric and
the Buchdahl parameter $k$ associated with it in pure $\mathcal{R}^{2}$
gravity \citep{Buchdahl-1962,Nguyen-2022-Buchdahl}, we carried the
concept over to the quadratic action. In a broader context, the Buchdahl
parameter $k$ should be a generic universal characteristic of a higher-derivative
theory of gravity. For the specific action $\mathcal{R}^{2}+\gamma\left(\mathcal{R}-2\Lambda\right)$,
we obtained a perturbative solution valid up to $\mathcal{O}\left(k^{2}\right)$.
The result is expressed by metric \eqref{eq:O-k-metric}--\eqref{eq:varphi-eqn}.
This metric possesses \emph{non-constant} scalar curvature induced
by a Buchdahl parameter $k$, in confirmation of our guiding intuition.\vskip4pt

To our knowledge, considerations of metrics with non-constant scalar
curvature have been exclusively in higher dimensions \citep{Ohashi-2015,Maeda-2010},
or in a generic $f(\mathcal{R})$ theory \citep{Zerbini-2010,Shankaranarayanan-2018,Odintsov-2020,Nashed-2019,Mazharimousavi-2011}.
In either situation, the generalized Lichnerowicz theorem as advocated
in \citep{Nelson-2010,Lu-2015-a,Lu-2015-b,Luest-2015-backholes} is
not applicable per se. The theorem, stated in Eq. \eqref{eq:Lichnerowicz},
was ``proved'' strictly for (i) the quadratic action \emph{and}
(ii) in $3+1$ dimensions. Our result, summed up in Lemma \ref{lem:lemma-1},
is thus \emph{novel}. It defeats the generalized Lichnerowicz theorem
which relied on overly strong restrictions assumed in the ``proofs'';
see Sec. \ref{sec:Motivation}. \vskip4pt

Considering the breakdown of the generalized Lichnerowicz theorem,
the $\mathcal{R}^{2}$ term should be reinstated in the full quadratic
action $\gamma\left(\mathcal{R}-2\Lambda\right)+\beta\,\mathcal{R}^{2}-\alpha\,\mathcal{C}^{\mu\nu\rho\sigma}\mathcal{C}_{\mu\nu\rho\sigma}$,
in counter to the practice adopted in \citep{Lu-2015-a,Lu-2015-b};
see our Sec. \ref{sec:Implications} herein for discussions. It remains
to be seen whether a perturbative metric, akin to metric \eqref{eq:O-k-metric}--\eqref{eq:varphi-eqn},
can be found for the full quadratic action, $\gamma\left(\mathcal{R}-2\Lambda\right)+\beta\,\mathcal{R}^{2}-\alpha\,\mathcal{C}^{\mu\nu\rho\sigma}\mathcal{C}_{\mu\nu\rho\sigma}$.
This would be an interesting possibility for future research.\vskip4pt

The solution \eqref{eq:O-k-metric}--\eqref{eq:varphi-eqn} obtained
herein should be applicable as long as $e^{k\,\varphi(r)}\approx1$,
which means at large distances. Close to the interior-exterior boundary
$r\approx r_{\text{s}}$, the solution should break down. The L\"u-Perkins-Pope-Stelle
ansatz could be suitable for the full quadratic action in the regions
close to the interior-exterior boundary. We touched upon the aspects
-- advantages and caveats -- of this direction in Sec. \ref{sec:Implications}.\vskip8pt

\emph{On the no-hair theorem:} As a member of the $f(\mathcal{R})$
family, the $\mathcal{R}+\mathcal{R}^{2}$ action is equivalent to
a scalar-tensor theory. Within scalar-tensor theories, Sotiriou and
Faraoni \citep{SotiriouFaraoni-2012} generalized Hawking's proof
\citep{Hawking-1972-BD} that outside of a horizon -- \emph{provided
that} there is one -- the scalar field must be constant (hence making
the Kerr-Newman metric an inevitable outcome of gravitational collapse).
In support of this proof, Agnese and La Camera \citep{Agnese-1995}
illustrated that the Campanelli-Lousto solution \citep{Campanelli-1993}
of Brans-Dicke gravity lacks a horizon; instead, it represents a naked
singularity or a wormhole, depending on whether $\gamma<1$ or $\gamma>1$.
These results indicate that a higher-derivative Buchdahl parameter,
facilitated by a relaxed boundary condition in the quadratic field
equation, could -- in qualified circumstances -- turn an $\mathcal{R}+\mathcal{R}^{2}$
spacetime structure into a naked singularity or a wormhole. Whether
this conclusion is applicable for the full quadratic action, $\gamma\left(\mathcal{R}-2\Lambda\right)+\beta\,\mathcal{R}^{2}-\alpha\,\mathcal{C}^{\mu\nu\rho\sigma}\mathcal{C}_{\mu\nu\rho\sigma}$,
is an open question. \vskip4pt
\begin{center}
-----------------$\infty$-----------------\vskip4pt
\par\end{center}

In closing, this paper is the third and final installment of our ``Beyond
Schwarzschild--de Sitter spacetimes'' series. The series started
by advancing the obscure six-decades-old Buchdahl program to attain
a novel exhaustive class of Buchdahl-inspired vacua for pure $\mathcal{R}^{2}$
gravity \citep{Nguyen-2022-Buchdahl}. It then progressed to a \emph{closed
analytical} metric describing an asymptotically flat non-Schwarzschild
spacetime with novel surprising properties \citep{Nguyen-2022-Lambda0}.
The series closes with a perturbative metric for the $\mathcal{R}^{2}+\mathcal{R}+\Lambda$
action, presented in this paper. The three sets of Buchdahl-inspired
spacetimes, uncovered in our three papers in sequel, reveal a host
of new interesting phenomenology that transcends the Einstein-Hilbert
paradigm.
\begin{acknowledgments}
I thank the anonymous referee of this current paper for pointing out
the latent role of the higher-derivative Buchdahl parameter in producing
naked singularities, which I addressed using Hawking's no-hair theorem
\citep{SotiriouFaraoni-2012,Agnese-1995,Campanelli-1993,Hawking-1972-BD}.
This paper was conceived partly by way of the anonymous referee of
my previous paper \citep{Nguyen-2022-Buchdahl} who motivated me to
strengthen the relation of my work to the generalized Lichnerowicz
theorem and, subsequently, the L\"u-Perkins-Pope-Stelle solution
in Einstein-Weyl gravity \citep{Nelson-2010,Lu-2015-a,Lu-2015-b,Luest-2015-backholes}.
I thank Dieter L\"ust and Tiberiu Harko for their encouragements.
The valuable help and technical insights of Richard Shurtleff are
acknowledged. I thank Sergei Odintsov and Timothy Clifton for their
supportive comments.
\end{acknowledgments}


\begin{thebibliography}{10}
\bibitem{Nelson-2010}W. Nelson, \textcolor{black}{\emph{Static solutions
for fourth order gravity}}\textcolor{black}{, Phys. Rev. D }\textbf{\textcolor{black}{82}}\textcolor{black}{,
104026 (2010), }\textcolor{purple}{\href{https://arxiv.org/abs/1010.3986}{arxiv:1010.3986 [gr-qc]}}

\bibitem{Lu-2015-a}H. L\"u, A. Perkins, C.N. Pope, and K.S. Stelle,
\emph{Black holes in higher-derivative gravity,} Phys. Rev. Lett.
\textbf{114}, 171601 (2015), \textcolor{purple}{\href{https://arxiv.org/abs/1502.01028}{arxiv:1502.01028 [hep-th]}}

\bibitem{Lu-2015-b}H. L\"u, A. Perkins, C.N. Pope, and K.S. Stelle,
\emph{Spherically symmetric solutions in higher-derivative gravity},
Phys. Rev. D \textbf{92}, 124019 (2015), \textcolor{purple}{\href{https://arxiv.org/abs/1508.00010}{arXiv:1508.00010 [hep-th]}}

\bibitem{Luest-2015-backholes}A. Kehagias, C. Kounnas, D. L\"ust,
and A. Riotto, \emph{Black hole solutions in $R^{2}$ gravity}, J.
High Energy Phys. \textbf{05} (2015) 143, \textcolor{purple}{\href{https://arxiv.org/abs/1502.04192}{arxiv:1502.04192 [hep-th]}}

\bibitem{Nguyen-2022-Buchdahl}H.K. Nguyen, \emph{Beyond Schwarzschild-de
Sitter spacetimes: I. A new exhaustive class of metrics inspired by
Buchdahl for pure $R^{2}$ gravity in a compact form, }Phys. Rev.
D \textbf{106}, 104004 (2022), \textcolor{purple}{\href{https://arxiv.org/abs/2211.01769}{https://arxiv.org/abs/2211.01769}}

\bibitem{Buchdahl-1962}H.A. Buchdahl, \emph{On the Gravitational
Field Equations Arising from the Square of the Gaussian Curvature},
Nuovo Cimento \textbf{23}, 141 (1962), \textcolor{purple}{\href{https://link.springer.com/article/10.1007/BF02733549}{https://link.springer.com/article/10.1007/BF02733549}}

\bibitem{Pravda-2018}J. Podolsk\'y, R. \v{S}varc, V. Pravda, and
A. Pravdov\'a, \emph{Explicit black hole solutions in higher-derivative
gravity}, Phys. Rev. D \textbf{98}, 021502 (2018), \textcolor{purple}{\href{https://arxiv.org/abs/1806.08209}{arXiv:1806.08209 [gr-qc]}}

\bibitem{Kokkotas-2017}K. Kokkotas, R. A. Konoplya, and A. Zhidenko,
\emph{Non-Schwarzschild black-hole metric in four dimensional higher
derivative gravity: Analytical approximation}, Phys. Rev. D \textbf{96},
064007 (2017), \textcolor{purple}{\href{https://arxiv.org/abs/1705.09875}{arXiv:1705.09875 [gr-qc]}}

\bibitem{Goldstein-2017}K. Goldstein and J.J. Mashiyane,\emph{ Ineffective
higher derivative black hole hair}, Phys. Rev. D \textbf{97}, 024015
(2018), \textcolor{purple}{\href{https://arxiv.org/abs/1703.02803}{arXiv:1703.02803 [hep-th]}}

\bibitem{Held-2022}A. Held and J. Zhang, \emph{Instability of spherically-symmetric
black holes in Quadratic Gravity}, Phys. Rev. D \textbf{107}, 064060
(2023),\emph{ }\textcolor{purple}{\href{https://arxiv.org/abs/2209.01867}{arXiv:2209.01867 [gr-qc]}}

\bibitem{Rinaldi-2020}C. Dioguardi and M. Rinaldi, \emph{A note on
the linear stability of black holes in quadratic gravity}, Eur. Phys.
J. Plus \textbf{135}, 920 (2020), \textcolor{purple}{\href{https://arxiv.org/abs/2007.11468}{arXiv:2007.11468 [gr-qc]}}

\bibitem{Bonanno-2019}A. Bonanno and S. Silveravalle, \emph{Characterizing
black hole metrics in quadratic gravity}, Phys. Rev. D \textbf{99},
101501 (2019), \textcolor{purple}{\href{https://arxiv.org/abs/1903.08759}{arXiv:1903.08759 [gr-qc]}}

\bibitem{Shurtleff-2022}R. Shurtleff, \textcolor{purple}{\href{http://www.wolframcloud.com/obj/shurtleffr/Published/20220401QuadraticGravityBuchdahl1.nb}{Link to notebook}}
(2022)

\bibitem{Nguyen-2022-Lambda0}H.K. Nguyen, \emph{Beyond Schwarzschild-de
Sitter spacetimes: II. An exact non-Schwarzschild metric in pure $R^{2}$
gravity and new anomalous properties of $R^{2}$ spacetime}, Phys.
Rev. D \textbf{107}, 104008 (2023), \textcolor{purple}{\href{https://arxiv.org/abs/2211.03542}{https://arxiv.org/abs/2211.03542}}

\bibitem{Shapiro-2004}D.F. Carneiro, E.A. Freiras, B. Goncalves,
A.G. de Lima, and I.L. Shapiro, \emph{On useful conformal transformations
in General Relativity}, Gravit. Cosmol. \textbf{10}, 305 (2004), \textcolor{purple}{\href{https://arxiv.org/abs/gr-qc/0412113}{arXiv:gr-qc/0412113}}

\bibitem{Stelle-1977}K.S. Stelle, \emph{Renormalization of higher-derivative
quantum gravity,} Phys. Rev. D \textbf{16}, 953 (1977)

\bibitem{Stelle-1978}K.S. Stelle, \emph{Classical Gravity with Higher
Derivatives}, Gen. Relativ. Gravit. \textbf{9}, 353 (1978)

\bibitem{Maeda-2010}H. Maeda, \emph{Gauss-Bonnet black holes with
nonconstant curvature horizons}, Phys. Rev. D \textbf{81}, 124007
(2010), \textcolor{purple}{\href{https://arxiv.org/abs/1004.0917}{arXiv:1004.0917 [gr-qc]}}

\bibitem{Ohashi-2015}S. Ohashi and M. Nozawa, \emph{Lovelock black
holes with non-constant curvature horizon}, Phys. Rev. D \textbf{92},
064020 (2015), \textcolor{purple}{\href{https://arxiv.org/abs/1507.04496}{arXiv:1507.04496 [gr-qc]}}

\bibitem{Zerbini-2010}L. Sebastiani and S. Zerbini, \emph{Static
spherically symmetric solutions in $F(R)$ gravity}, Eur. Phys. J.
C \textbf{71}, 1591 (2011), \textcolor{purple}{\href{https://arxiv.org/abs/1012.5230}{arXiv:1012.5230 [gr-qc]}}

\bibitem{Mazharimousavi-2011}S.H. Mazharimousavi, M. Halilsoy, and
T. Tahamtan, \emph{Solutions for $f(R)$ gravity coupled with electromagnetic
field}, Eur. Phys. J. C \textbf{72}, 1851 (2012), \textcolor{purple}{\href{https://arxiv.org/abs/1110.5085}{arXiv:1110.5085 [gr-qc]}}

\bibitem{Nashed-2019}G.G.L. Nashed and S. Capozziello, \emph{Charged
spherically symmetric black holes in $f(R)$ gravity and their stability
analysis}, Phys. Rev. D \textbf{99}, 104018 (2019), \textcolor{purple}{\href{https://arxiv.org/abs/1902.06783}{arXiv:1902.06783 [gr-qc]}}

\bibitem{Odintsov-2020}E. Elizalde, G.G.L. Nashed, S. Nojiri, and
S.D. Odintsov, \emph{Spherically symmetric black holes with electric
and magnetic charge in extended gravity: physical properties, causal
structure, and stability analysis in Einstein\textquoteright s and
Jordan\textquoteright s frame}, Eur. Phys. J. C \textbf{80}, 109 (2020),
\textcolor{purple}{\href{https://arxiv.org/abs/2001.11357}{arXiv:2001.11357 [gr-qc]}}

\bibitem{Shankaranarayanan-2018}S. Xavier, J. Mathew, and S. Shankaranarayanan,
\emph{Infinitely degenerate exact Ricci-flat solutions in $f(R)$
gravity}, Class. Quant. Grav. \textbf{37}, 225006 (2020), \textcolor{purple}{\href{https://arxiv.org/abs/2003.05139}{arXiv:2003.05139 [gr-qc]}}

\bibitem{SotiriouFaraoni-2012}T.P. Sotiriou and V. Faraoni, \emph{Black
holes in scalar-tensor gravity}, Phys. Rev. Lett. \textbf{108}, 081103
(2012), \textcolor{purple}{\href{https://arxiv.org/abs/1109.6324}{arXiv:1109.6324 [gr-qc]}}

\bibitem{Hawking-1972-BD}S.W. Hawking, \emph{Black holes in Brans-Dicke
theory of gravitation}, Commun. Math. Phys. \textbf{25}, 167 (1972)

\bibitem{Agnese-1995}A.G. Agnese and M. La Camera, \emph{Wormholes
in the Brans-Dicke theory of gravitation}, Phys. Rev. D \textbf{51},
2011 (1995)

\bibitem{Campanelli-1993}M. Campanelli and C. Lousto, \emph{Are Black
Holes in Brans-Dicke Theory precisely the same as in General Relativity?},
Int. J. Mod. Phys. D \textbf{2}, 451 (1993), \textcolor{purple}{\href{https://arxiv.org/abs/gr-qc/9301013}{arXiv:gr-qc/9301013}}
\end{thebibliography}
\end{document}